# First Field-Trial Demonstration of L4 Autonomous Optical Network for Distributed AI Training Communication: An LLM-Powered Multi-AI-Agent Solution

Yihao Zhang, Qizhi Qiu, Xiaomin Liu, Dianxuan Fu, Xingyu Liu, Leyan Fei, Yuming Cheng, Lilin Yi, Weisheng Hu, and Qunbi Zhuge*

State Key Laboratory of Photonics and Communications, Shanghai Jiao Tong University, Shanghai 200240, China. *qunbi.zhuge@sjtu.edu.cn

**Abstract** *We demonstrate the first cross-domain cross-layer level-4 autonomous optical network via a multi-AI-agent system. Field trials show ~98% task completion rate across the distributed AI training lifecycle—3.2× higher than single agents using advanced LLMs.* 

## Introduction

The explosive development of network services such as distributed training for large artificial intelligence (AI) models is reshaping mega data centers to geo-distributed architectures interconnected via optical networks [1]. Since collaborative resource utilization across distributed facilities is essential for training workloads, this evolution introduces significant complexity in network management, as controllers must operate across multiple domains, spanning from intra- and inter-datacenters to long-haul wide area networks. Moreover, distributed training imposes stringent reliability requirements as it should restart from a checkpoint if a failure happens [2]. Therefore, in terms of distributed training communications, resilient operations and rapid fault recovery are essential. In this context, pursuing level-4 (L4) network automation is crucial, which is defined to enable zero-wait, zero-touch, zero-trouble management in a complex cross-domain environment [3].

Concurrently, the emergence of AI agents—defined as entities that perceive, reason, and act through large language models (LLMs)—has demonstrated cutting-edge breakthrough in complex decision-making and coordination tasks [4]. For cross-domain network control, a single AI agent faces inherent limitations due to the information isolation and controller incompatibility among different domains. Therefore, a multi-agent paradigm becomes necessary, where distinctive AI agents manage specific domains and interact through natural language. To date, the multi-agent solution for multi-domain optical networks has not been investigated.

In this paper, we present the first field-trial demonstration of an L4 autonomous optical network, realized through our LLM-powered multi-agent system AutoLight. With the proposed Chain-of-Identity technique, AutoLight enables unified autonomous management across heterogeneous scenarios integrating long-haul transmissions, data center interconnection (DCI), and intra-datacenter networks. We implement AutoLight to an emulated lifecycle for distributed training communication through comprehensive experiments across all above scenarios, and across both physical and network layers. During the lifecycle, AutoLight achieves ~98% task completion rate, which is ~3.2× of that achieved by single AI agents empowered by the advanced LLMs on identical tasks.

## L4 Autonomous Network Demonstration

### *Demonstration Setup*

To demonstrate an L4 network automation, we develop an integrated validation platform that emulates a geo-distributed data center cluster. This setup connects two DCI metro networks via a long-haul backbone link spanning two distinct domains, as illustrated in Fig. 1(a).

The *backbone* part depicted in inset **(i)** is a 440-km field-deployed testbed from Shanghai to Hangzhou in China, as shown in Fig. 1(b). This 4-span transmission link is controlled and monitored by two isolated domains. Each domain consists of two G.652D fibers and three commercial C-band EDFAs. The signal power can be monitored at both input and output of these EDFAs. Six 400/200Gbps transponders operating at 63.9 GBaud are utilized for transmission. Combined with the dummy signals generated by an ASE noise source and a programmable filter, a 30-wavelength transmission is demonstrated. Hardware control is implemented through NETCONF protocol with YANG models.

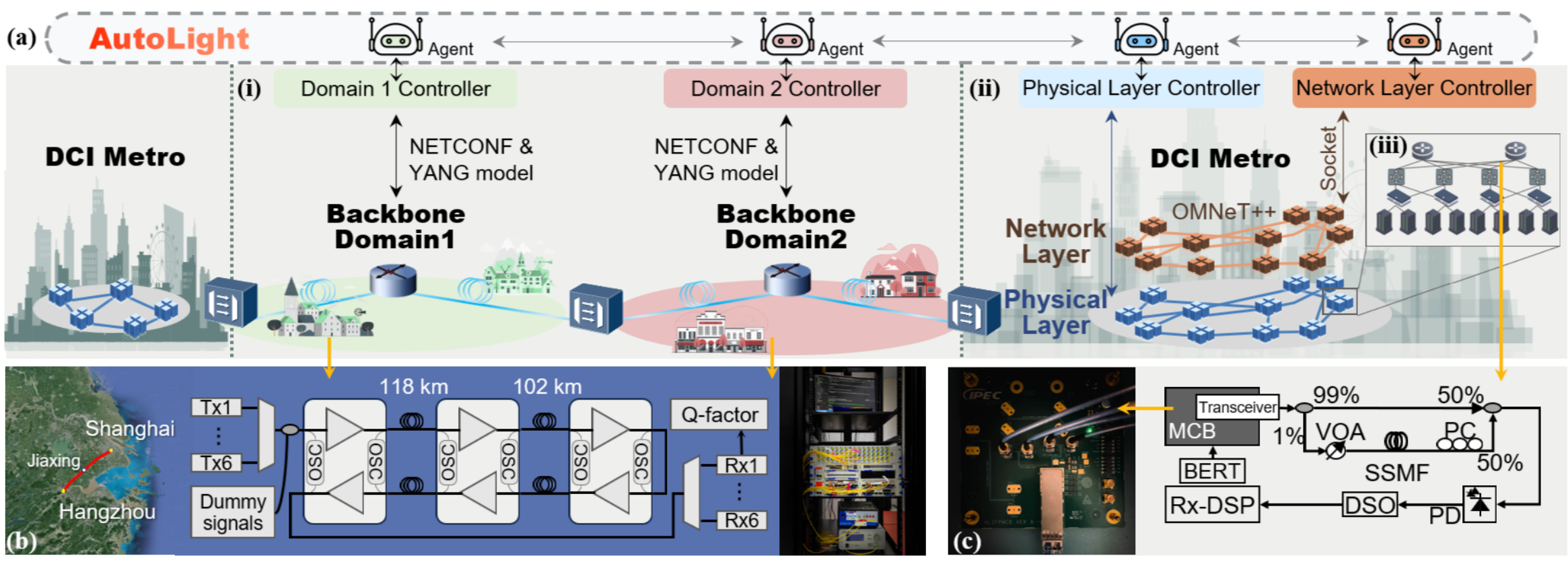

**Fig. 1:** Setup of the demonstration: **(a)** schematics, the **(b)** long-haul and **(c)** IMDD system.

The *DCI metro network* implementation in inset **(ii)** adopts a topology with 14 nodes [5]. Within each node site of this DCI metro network, an *intra-datacenter network* comprising 8 server groups is simulated as shown in inset **(iii)**, with each group containing 8 servers interconnected through a two-tier spine-leaf architecture [6]. The corresponding network layer is emulated by OMNeT++ [7], which communicates with the controller via a socket-based interface. For the physical layer implementation, an IMDD system is constructed to emulate the optical links in the intra-datacenter network, as illustrated in Fig. 1(c). A 53Gbps PAM-4 optical signal is generated by an optical transceiver. To evaluate system resilience under impairments, we introduced multipath interference (MPI) through delayed reflection path.

***Emulated Lifecycle for Distributed Training Communication***

A lifecycle for distributed AI training communication is emulated. Initially, the computing resource needs for the distributed training workload are assessed. Before the training starts, demand matrices are generated for each training epoch, which guide the physical- and network-layer resource allocation during training. When local DCI computing resources are insufficient, spectrum resources are requested from the backbone network to connect the secondary DCI with new wavelengths. Meanwhile, continuous physical-layer failure detection is performed. Upon detection, the system classifies and localizes the failure, and implements proactive rerouting to avoid training disruption.

For systematic evaluation, we decompose this lifecycle into four tasks. *Task1—real-time resource allocation* illustrates the DCI conducting resource allocation during distributed training execution at each epoch. *Task2—backbone wavelength establishment* demonstrates the backbone network being requested to establish the inter-DCI transmission. *Task3—DCI failure management* emulates the detection and mitigation for physical-layer failures in the DCI metro network. *Task4—backbone failure management* emulates coordinated troubleshooting of failures in two information-isolated backbone domains.

**AutoLight: LLM-Powered Multi-Agent System**

AutoLight employs a hierarchical multi-agent architecture, where multiple ReAct agents [8] collaborate through structured interactions. Specifically, it comprises two primary categories of ReAct agents: Planner agents and Task agents, as illustrated in Fig. 2(a). Each ReAct agent is powered by an LLM with tools. Planner agents serve as high-level coordinators responsible for problem decomposition and execution orchestration. They maintain a *plan tracking table* which records task progress and orchestrate next actions across the multi-agent system. The plan tracking table can ensure continuous operation of tasks, preventing anomalous task termination. While Planner agents focus on coordination and are bound with minimal toolsets for basic operations, Task agents are specialized for executing complex operations. Each Task agent is equipped with a comprehensive toolkit tailored to its designated function, whether it be resource allocation, digital twin, or failure handling. As one of these Task agents, the Knowledge Retriever employs retrieval-augmented generation (RAG) to access information from external documents, ensuring operation reliability [10]. The detailed structure of AutoLight is depicted in Fig. 2(b).

For effective interaction among agents, we propose a novel technique called Chain of Identity (CoI). As shown in Fig. 2(a), CoI includes three mechanisms. (1) *Formatted handoffs:* agents utilize structured texts for inter-agent transfers. The handoff consists of a greeting that explicitly identifies the target agent, a query, and a parameter section containing essential arguments. (2) *Pseudo-SystemMessage injection*: handoff transfers are implemented by designed tools that embed system-level instructions within ToolMessages. These instructions contain the identity and core responsibility of the target agent. (3) *Pre-execution declaration*: before any action, agents are required to issue an identity declaration and a verification of received ToolMessage with handoffs and pseudo-SystemMessages. Through CoI, continuous identity awareness and contextual consistency across agent interactions are ensured.

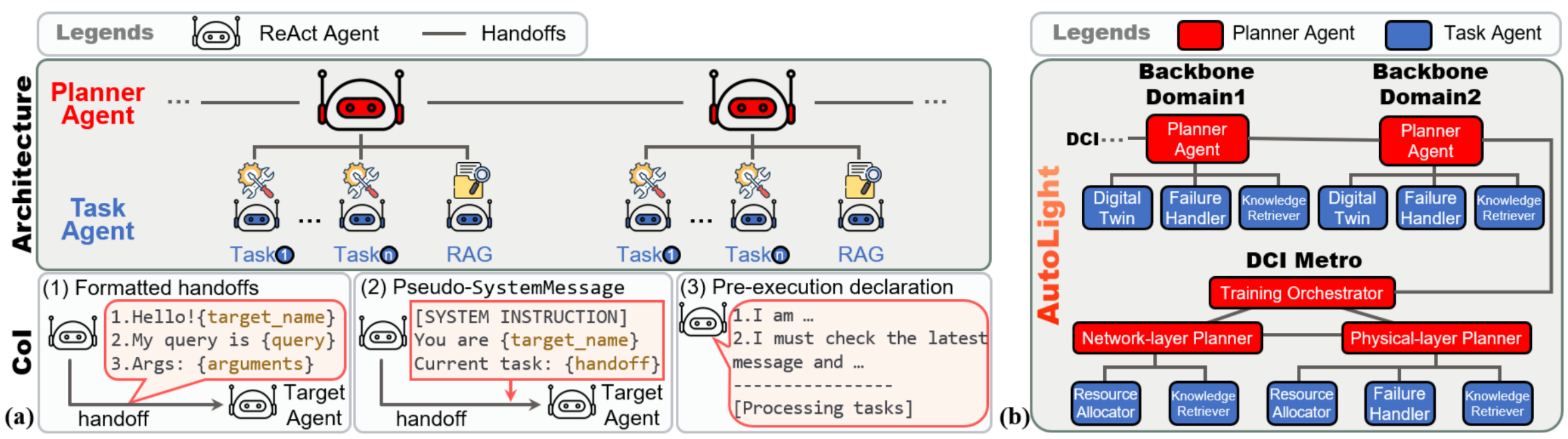

**Fig. 2:** The framework diagram of AutoLight: **(a)** overall schematic and **(b)** detailed structure. AutoLight is developed based on LangGraph [9].

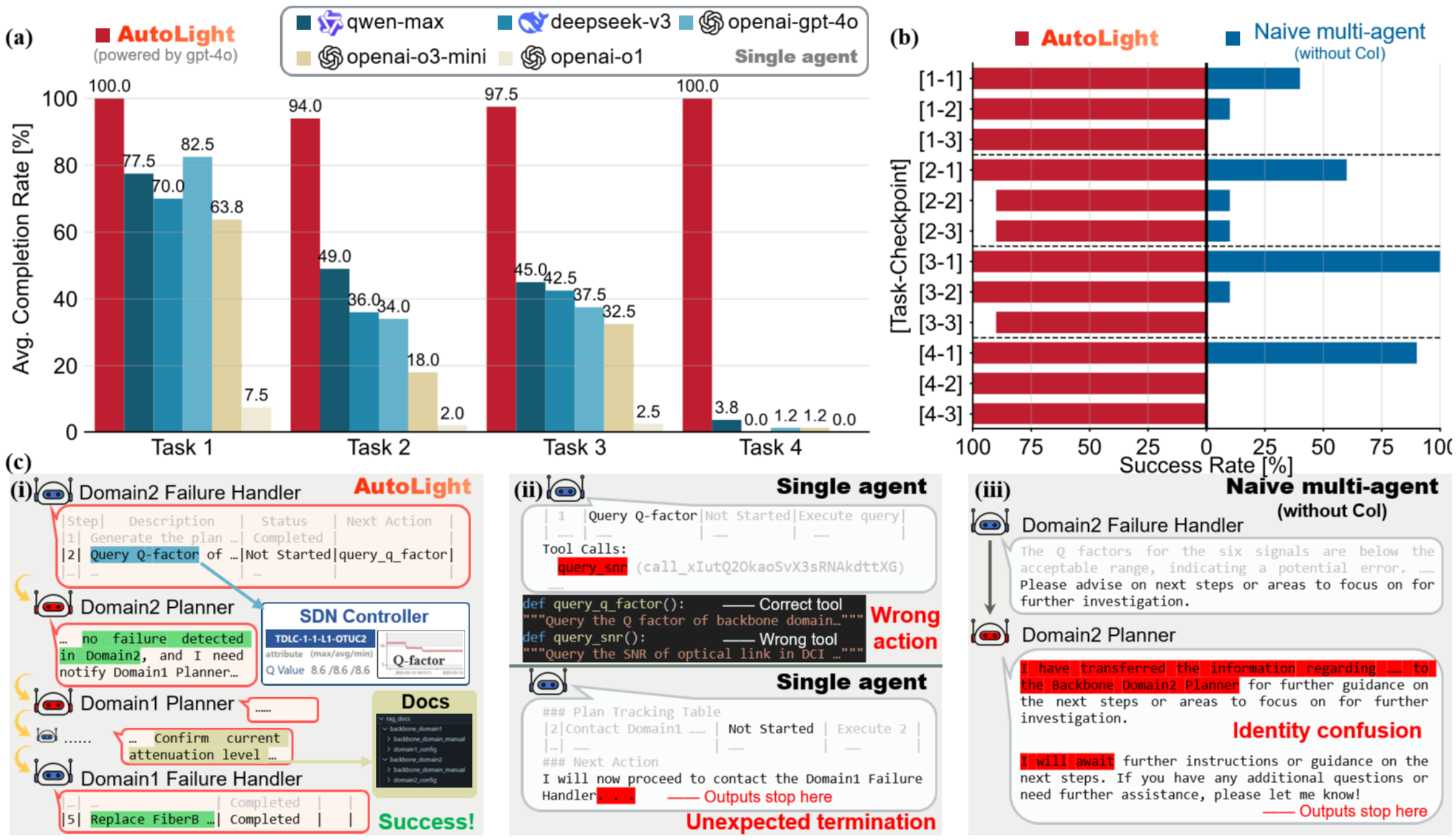

**Fig. 3: (a)** Comparison of average task completion rate between AutoLight and single agents. **(b)** Comparison of success rates at different checkpoints between AutoLight and the naive multi-agent. **(c)** AutoLight autonomously handles Task4, while the baseline schemes fail to handle.

## Results [1]

In our demonstration, AutoLight achieves autonomous management across all the four tasks, with experiments conducted over a total of 40 trials (10 trials per task). To evaluate its advantages, AutoLight is compared with two baselines: 1) a single agent for all domains and 2) a naive multi-agent without the proposed CoI. For the single agent using advanced LLMs with all necessary tools and cross-domain information, the average completion rates are presented in Fig. 3(a). Despite utilizing the advanced LLMs, single agents achieve only ~30% average completion rate across these tasks, while AutoLight (w/ GPT-4o) reaches ~98% completion rate, representing a ~3.2× improvement. In Fig. 3(b), AutoLight is compared to the naive multi-agent without CoI, where the average success rates for specific checkpoints of each task are illustrated. Results show that the naive multi-agent completes only a small fraction of checkpoints, whereas AutoLight successfully completes almost all checkpoints, demonstrating superior robustness.

In Fig. 3(c), we select Task4 as an example to elucidate the underlying reasons for AutoLight's superiority. As shown in the inset **(i)**, AutoLight's agents collaborate effectively, precisely perform planning, controller querying, document retrieval, and finally localize the aging fiber. In contrast, the single agent depicted in **(ii)** fails to manage the overwhelming volume of information and tools for all domains, leading to disorientation during task execution. Moreover, the single agent may terminate unexpectedly after generating lengthy outputs or performing numerous operations, failing to complete the task. For the naive multi-agent, as shown in **(iii)**, identity confusion often occurs for each agent after transferring without CoI, which causes unexpected terminations as well.

To conclude, AutoLight shows superior capabilities in cross-domain scenarios, meeting all the criteria defined in [3]. AutoLight thereby successfully achieves L4 autonomous optical networks.

## Conclusions

We present the first field-trial demonstration of an L4 autonomous optical network operation through AutoLight, an LLM-powered multi-agent system across multiple domains and layers. The demonstration encompasses both long-haul and short-reach transmission, as well as a metro network OMNeT++ platform, validating AutoLight's robustness across diverse scenarios. Through the novel hierarchical structure and CoI, AutoLight achieves a ~98% task accurate-completion rate, which is ~3.2× higher than advanced LLM-enabled single-agent approaches.

[1] Full description of all tasks and evaluation details can be found at: https://github.com/AutoLight2025/AutoLight .

## Acknowledgements

This work was supported by Shanghai Pilot Program for Basic Research - Shanghai Jiao Tong University (21TQ1400213) and National Natural Science Foundation of China (62175145).